\documentclass[runningheads]{llncs}
\usepackage[T1]{fontenc}
\usepackage{graphicx}
\usepackage{thm-restate}
\usepackage[rflt]{floatflt}
\usepackage{caption}
\usepackage{booktabs} 
\usepackage{mathtools}
\usepackage{fullpage}
\usepackage{xspace}
\usepackage{soul}
\usepackage{amsfonts}
\usepackage{setspace}
\usepackage{extpfeil}
\usepackage{xcolor}
\usepackage{macros}
\usepackage{tikz}
\usepackage{hyperref}
\usepackage{cleveref}

\renewenvironment{proof}{{\textit{Proof. }}}{\qed}
\definecolor{MNBlue}{rgb}{0.1, 0.1, 0.44}
\usepackage[notion,quotation,xcolor,makeidx,paper]{knowledge}
\hypersetup{breaklinks=true,colorlinks=true,
linkcolor=MNBlue,urlcolor=blue,citecolor=blue, 
bookmarks=true,bookmarksopenlevel=2}
\knowledge{notion}
 | \dwroca
 | \dwrocas
\knowledge{notion}
 | field
\knowledge{notion}
 | pumping
\knowledge{notion}
 | \fst
\knowledge{notion}
 | \snd
 \knowledge{notion}
 | word
\knowledge{notion}
 | \we
 | weight-effect
\knowledge{notion}
 | \ce
 | counter-effect
\knowledge{notion}
 | efficiency
\knowledge{notion}
 | simple cycle
\knowledge{notion}
 | \equiv
 |\not\equiv
\knowledge{notion}
 | minimal witness
\knowledge{notion}
 | \uwa \Autom
 | \uwa {\Autom_1}
 | \uwa {\Autom_2}
 | underlying uninitialised weighted automaton
 \knowledge{notion}
 | \Dcwak \Autom \Butom k
 | \Dcwak{\Autom}{\Butom}{k}
 | \Dcwak{\Autom_i}{\uwa{\Autom_j}}{\K}
 \knowledge{notion}
 | \Dwak \Autom \Butom k
 \knowledge{notion}
 | \K
  \knowledge{notion}
 | \Pthree
  \knowledge{notion}
 | \Pzero
   \knowledge{notion}
 | \Pone
   \knowledge{notion}
 | \Ptwo
   \knowledge{notion}
 | \mrun
  \knowledge{notion}
 | \sim
 | \not\sim
    \knowledge{notion}
 | \aw
     \knowledge{notion}
 | \NWA
     \knowledge{notion}
 | \WA
   \knowledge{notion}
 | \dist
     \knowledge{notion}
 | surely-equivalent
     \knowledge{notion}
 | surely-nonequivalent
   \knowledge{notion}
 | unresolved
      \knowledge{notion}
 | initial space
     \knowledge{notion}
 | belt space
   \knowledge{notion}
 | background space
    \knowledge{notion}
 | \rep

\begin{document}
\title{Equivalence of Deterministic Weighted Real-time One-Counter Automata}
\titlerunning{Equivalence of DWROCA}
%

 \author{%
     Prince Mathew \inst{1} \and
     Vincent Penelle\inst{2} \and
     Prakash Saivasan\inst{3} \and
     A. V. Sreejith\inst{4} 
 }
%
\authorrunning{P. Mathew et al.}
  \institute{
      Indian Institute of Technology Goa,
      \email{prince@iitgoa.ac.in}\\
       Univ. Bordeaux, CNRS,  Bordeaux INP, LaBRI, UMR 5800, F-33400, Talence, France,
      \email{vincent.penelle@u-bordeaux.fr}\\
       The Institute of Mathematical Sciences, HBNI,
      \email{prakashs@imsc.res.in}\\
      Indian Institute of Technology Goa,
      \email{sreejithav@iitgoa.ac.in}
}
\maketitle              
\setcounter{tocdepth}{3}


\begin{abstract} 
This paper introduces deterministic weighted real-time one-counter automaton ("\dwroca"). A \dwroca is a deterministic real-time one-counter automaton whose transitions are assigned a weight from a field. Two "\dwrocas" are equivalent if 
every word accepted by one is accepted by the other with the same weight. 
"\dwroca" is a sub-class of weighted one-counter automata with counter-determinacy. It is known that the equivalence problem for this model is in $\CF{P}$~\cite{wodca}. This paper gives a simpler proof and a better polynomial-time algorithm for checking the equivalence of two "\dwrocas".

\end{abstract}


\section{Introduction}
\label{sec:introduction}
In this work, we give a polynomial-time algorithm to check the equivalence of deterministic weighted real-time one-counter automata ("\dwroca") where the weights are from a "field" (possibly infinite). These are weighted one-counter automata with the ``deterministic'' condition ---  at most one transition on a letter from a state, and the counter is modified by at most $one$ on a transition. Hence, for a word over a finite alphabet, there is at most one run starting from the initial configuration that determines its accepting weight.
Additionally, ``real-time'' indicates that there are no epsilon transitions in this system.

We say that two "\dwrocas" are equivalent if any word accepted by one of the automata is accepted by the other with the same weight. Consider two "\dwrocas" that are not equivalent and let $w$ be a word of minimal length that distinguishes the two machines. It suffices to show that there is a polynomial bound on the length of $w$. Two weighted automata that ``simulate'' the run of the "\dwrocas" up to that bound can then be used to check equivalence. 
To prove a bound on the length of $w$, we give a polynomial that bounds the counter values during the run of $w$ on both machines. Our proof strategy borrows the ``belt technique'' developed in the context of deterministic real-time one-counter automata by B\"ohm et al.~\cite{droca} (also see~\cite{bisim,bisimjour}), which is inspired by Valiant and Paterson's work~\cite{doca}.  We introduce a \emph{"pumping"} technique (see \Cref{generalTheorem}) to adapt their proof strategy for checking the equivalence of two "\dwrocas".

\subsection{Related Work}
 One-counter automata are pushdown automata with a singleton stack alphabet. 
 It is well known that the equivalence problem for non-deterministic pushdown automata is undecidable. On the other hand, S\'enizergues~\cite{dpdaequiv} proved that it is decidable for deterministic pushdown automata. It was later proved that there is a primitive recursive algorithm for this~\cite{dpdalowerbound}. 
The equivalence problem for deterministic one-counter automata is $\CF{NL}$-complete~\cite{docaequiv}. Studies on probabilistic pushdown automata conducted by
Forejt et al.~\cite{openpda} showed that the equivalence problem of probabilistic pushdown automata is equivalent to the multiplicity equivalence of context-free grammars. The latter problem is not known to be decidable. The decidability of equivalence is known for some special sub-classes of probabilistic pushdown automata. 
The equivalence problem is at least as hard as polynomial identity testing~\cite{openpda} even when the alphabet contains only one letter. For the case of visibly probabilistic pushdown automata, there is a randomised polynomial time algorithm for checking equivalence~\cite{Kiefer}. However, there is a polynomial time reduction from polynomial identity testing to this problem and is hence not likely to be in $\CF{P}$.
In a recent result, Mathew et al.~\cite{wodca} showed that the equivalence problem for weighted one-deterministic counter automata, which is a sub-class of weighted one-counter automata over fields, is decidable in polynomial time.

\Cref{sec:prelims} contains the basic definitions that are used in the paper. \Cref{sec:ptime} gives a polynomial time algorithm to check the equivalence of two "\dwrocas" given the existence of a polynomial length word that distinguishes them and the existence of such a word is proved in \Cref{sec:polywitness}. 
\Cref{sec:conclusion} gives a short conclusion.


\section{Preliminaries}
\label{sec:prelims}

We use the symbol $\N$ to denote the set $\{0,1,2,3,\ldots\}$.
A group is a triple $(S, \circ, e)$ where $S$ is a non-empty set, $\circ$ is an associative binary operation, and $e \in S$ is an identity element. The set $S$ is closed under the operation $\circ$ and for all $s\in S$, $e\circ s=s\circ e=s$. 
For all $s\in S$ there exists $s^{-1}\in S$ such that $s \circ s^{-1}=s^{-1} \circ s= e$. 

\AP
A ""field"" $\Sring = (S,+,\times,0_e,1_e)$ is a set $S$ with two commutative operations $+$ and $\times$ and distinguished elements $0_e$ and $1_e$ such that
\begin{itemize}
\item $(S, + , 0_e)$ and $(S\setminus\{0_e\}, \times, 1_e)$ are groups.
\item $\times$ distributes over $+$ (i.e., for all $ s,t,r\in S$, $s \times (t+r) = s\times t + s \times r$).
\end{itemize}
The operator $+$ is never used in the rest of the paper. We can assume that the weights come from a group. 

An alphabet is a finite non-empty set of letters. In this paper, we denote the alphabet by $\Sigma$. 
We use $\Sigma^*$ to denote the set of finite length words over $\Sigma$, and for all $l \in \N$, we use $\Sigma^{\leq l}$ (resp. $\Sigma^l$) to denote the set of words over $\Sigma$ having length less than or equal to $l$ (resp. exactly equal to $l$). Given a word $w\in\Sigma^*$, we use $|w|$ to denote the length of the word $w$.
We use the notation $[i,j]$ to denote the interval $\{i, i+1, \ldots , j\}$.
Given a word $w = w_0\cdots w_n$, we write $w_{[i\cdots j]}$ to denote the factor $w_i\cdots w_j$.
Given a tuple $e=(e_1,e_2)$, we use $e|_1$ to denote $e_1$ and $e|_2$ to denote $e_2$. 

\subsection{Deterministic weighted real-time one-counter automata}
\AP
{\sloppy
A deterministic weighted real-time one-counter automaton (""\dwroca"") over a "field" $\Sring$ is a tuple $\Autom=(Q, \Sigma, q_0,s_0,\delta_0,\delta_1,\eta_F)$, where
 \begin{itemize}
 \item[$-$] $Q$ is a finite, nonempty set of states.
 \item[$-$]  $\Sigma$ is the input alphabet.
 \item[$-$] $q_0 \in Q$ is the initial state.
 \item[$-$] $s_0 \in \Sring$ is the initial weight.
  \item[$-$]  $\delta_0 : Q \times \Sigma \to Q \times \{0,+1\} \times \Sring$  and $\delta_1 : Q \times \Sigma \to Q \times \{-1,0,+1\} \times \Sring$ are the transition functions. 
 \item[$-$]  $\eta_F: Q \to \Sring$ is a final distribution, that assigns an output weight to each state.
 \end{itemize}
}

We use $|\Autom|$ to denote the size of $\Autom$, which we consider to be $|Q|$. 
A \emph{configuration} of a "\dwroca" is a triple $c = \configoca{q_c}{n_c}{t_c}$, with $q_c \in Q$, $n_c \in \N$ and $t_c \in \Sring$.  The configuration $\configoca{q_0}{0}{s_0} \in Q\times \N \times \Sring$ is called the \emph{initial configuration} of $\Autom$. A \emph{transition} is a sextuple $\tau = ({p_\tau},d_\tau,a_\tau,\ce_\tau,s_\tau,{q_\tau})$ where ${p_\tau},{q_\tau}\in Q$ are states, $d_\tau \in \{0,1\}$ specifies which among $\delta_0$ and $\delta_1$ is used, $a_\tau\in \Sigma$, $\ce_\tau \in \{-1,0,1\}$ is the \emph{counter-effect}, and $s_\tau \in \Sring$ such that $\delta_{d_\tau}({p_\tau},a_\tau) = ({q_\tau},\ce_\tau,s_\tau)$. Intuitively, $\delta_0$ is used when you have a zero-test, and $\delta_1$ otherwise.
Given a transition $\tau$ and a configuration $c=\configoca{q_c}{n_c}{s_c}$, we denote the application of $\tau$ to $c$ as $\tau(c) = ({q_\tau},n_c+\ce_\tau,s_c\times s_\tau)$ if $q_c = {p_\tau}$ and $d_\tau = 0$ if and only if $n_c = 0$, and is undefined otherwise. Note that the counter values always stay positive, implying that you cannot perform a decrement operation on the counter from a configuration with counter value zero. 

\AP
Given a sequence of transitions $T = \tau_0\cdots \tau_{\ell-1}$, we denote $""word""(T) = a_{\tau_0}\cdots a_{\tau_{\ell-1}}$ the word labelling it, $""\we""(T) = s_{\tau_0} \times \cdots \times s_{\tau_{\ell-1}}$ its "weight-effect", and $""\ce""(T) = \ce_{\tau_0} + \cdots + \ce_{\tau_{\ell-1}}$ its "counter-effect". For all $0 \leq i < j \leq |\ell-1|$, we use $T_{i\cdots j}$ to denote the sequence of transitions $\tau_i\cdots \tau_j$ and $|T|$ to denote $\ell$. 

{
We call a sequence of transitions \emph{grounded} if there is an $i$ such that $d_{\tau_i} = 0$ and \emph{floating} otherwise.
We denote $\min_{\ce}(T) = \min_i("\ce"(\tau_0\cdots \tau_i))$ the minimal counter-effect of its prefixes and $\max_{\ce}(T) = \max_i("\ce"(\tau_0\cdots \tau_i))$ the maximal counter-effect of its prefixes.
}
We say that the sequence of transitions $T$ is \emph{valid} if for every $i\in[0,\ell-2]$, ${q_{\tau_i}} = {p_{\tau_{i+1}}}$. We will only consider valid sequences of transitions.

A \emph{run} $\pi$ is an alternate sequence of configurations and transitions denoted as $\pi = c_0 \tau_0 c_1 \cdots \tau_{\ell-1} c_\ell$ such that for every $i$, $c_{i+1} = \tau_i(c_i)$.
Given a sequence of transitions $T$ and a configuration $c$, we denote $T(c)$ the run obtained by applying $T$ to $c$ sequentially (if it is defined). The word labelling it, its length, "weight-effect" and "counter-effect" are those of its underlying sequence of transitions.
A \emph{sub-run} is a (syntactic) sub-word of a run that is also a run.
\AP
The run $\pi= c_1\tau_1c_2 \tau_2 \ldots c_i$ is called a ""simple cycle"", if  $i\leq|\Autom|$, $q_{c_1}=q_{c_i}$ and for all $j,k\in[1,i-1]$, $q_{c_j}=q_{c_k}$ if and only if $j=k$. 
 The ""efficiency"" of a "simple cycle"  $\varepsilon=\frac{-\ce(\pi)}{|\pi|}$, is the counter loss per length of the cycle.

Observe that, for a valid floating sequence of transitions, $T(c)$ is defined if and only if $n_c > - \min_\ce(T)$, and for a valid grounded sequence of transition, $T(c)$ is defined if and only if $n_c = - \min_\ce(T)$ and for every $i$, $d_{\tau_i} = 0$ if and only if $"\ce"(\tau_0\cdots \tau_{i-1}) = \min_\ce(T)$.  
In particular, observe that if a valid floating sequence of transitions $T$ is applicable to a configuration $\configoca qnt$, then for every $n^\prime \geq n$ and weight $s^\prime \in \Sring$, it is applicable to $\configoca{q}{n^\prime}{s^\prime}$.

Since the machine is deterministic, for any word $w$, there is at most one run labelled by $w$ starting in a given configuration $c_0$. We denote this run $\pi(w,c_0)$. 
A run $\pi$ is called an \emph{execution} if $c_0=\configoca{q_0}{0}{s_0}$ is the initial configuration of $\Autom$. We use $\pi(w)$ to denote the execution labelled by $w$. 
A run $\pi(w,c_0)=c_0 \tau_0 c_1 \cdots \tau_{\ell-1} c_\ell$ is also represented as  $c_0 \xrightarrow{w}c_\ell$. 
We use the notation $c_0 \rightarrow^* c_\ell$ to denote the existence of some word $w$ such that $c_0 \xrightarrow{w}c_\ell$.
The weight with which a word $w$ is accepted by $\Autom$ along the run $\pi(w,c_0)$ is denoted by $f_\Autom(w,c_0)= s_{c_0} \times \we(\pi(w,c_0))\times\eta_F(q_{c_\ell})$. We use the notation $f_\Autom(w)$ to denote $f_\Autom(w,{\configoca{q_0}{0}{s_0}})$. 

\AP
Let $\Autom$ and $\Butom$ be two \dwrocas. Consider the configurations $c_1$ of $\Autom$ and $c_2$ of $\Butom$. We say that 
$c_1""\equiv""_l c_2$ if and only if 
for all $w\in\Sigma^{\leq l}$,  $f_\Autom(w, c_1)= f_\Butom(w, c_2)$. 
 We say that the configurations  $c_1$ and $c_2$ are equivalent if and only if  $c_1 "\equiv"_l c_2$ for all $l\in \N$, and we denote this by  $c_1\equiv c_2$.
We say that $\Autom$ and $\Butom$ are equivalent if for all $w\in\Sigma^*$,  $f_\Autom(w)= f_\Butom(w)$.
A word $w\in\Sigma^*$ is called a \emph{non-equivalent witness} (or simply a $witness$) {for} \Autom and \Butom if and only if $f_\Autom(w)\neq f_\Butom(w)$. A \emph{""minimal witness""} is a non-equivalent witness with the minimal length.

Deterministic weighted automata (\dwa) is a restricted form of a "\dwroca" where there are no counter operations. The above definitions for "\dwroca" are also used for \dwa. 


Given a word $w = w_0\cdots w_n$, a configuration $c$, and a list  $I$  of disjoint intervals included in $[0,n]$ : $I = [i_0,j_0],[i_1,j_1],\cdots,[i_k,j_k]$, we call $w_I$ the word obtained by removing from $w$ all $w_\ell$ with $\ell \in I$.
\AP
A list of disjoint intervals $I$ is a \emph{""pumping""} of $w$ from $c$, if in $\pi(w,c)$ all sub-runs starting at position $i_d$ and ending in $j_d$ for some $d\in[0,k]$ are \emph{loops}, i.e., start and end in the same state, the minimal "counter-effect" of the obtained run is not smaller than the original one, i.e., $\min_{\ce}(\pi(w_I,c)) \geq \min_{\ce}(\pi(w,c))$ and the "pumping" does not introduce any new zero-tests, while still preserving the last zero-test. Informally, a \emph{pumping} is obtained by removing loops in a run while not decreasing the minimum counter value reached and without introducing additional zero-tests while still preserving the last zero-test.

\begin{theorem}\label{generalTheorem}
Given a word $w$ and two configurations $c$ and $c^\prime$, and two sequences of interval $I$ and $J$ such that:
\begin{itemize}
   \item $I$ and $J$ are disjoint.
   \item $I$ and $J$ are "pumping"s of $w$ from both $c$ and $c^\prime$.
   \item $f(w,c) \neq f(w,c')$.
\end{itemize}
We get that $I\uplus J$ is also a "pumping" of $w$ from both $c$ and $c^\prime$, and either $f(w_I,c) \neq f(w_I,c^\prime)$, or $f(w_J,c) \neq f(w_J,c^\prime)$ or $f(w_{I\uplus J},c) \neq f(w_{I \uplus J},c^\prime)$.
\end{theorem}

\begin{proof}
As $I$ and $J$ are disjoint, it is immediate that $I \uplus J$ is a "pumping" of $w$ from both configurations as the loops removed are disjoint. 
For the second point, we consider the weights of the transitions removed on the runs from $c$ (resp. $c^\prime$) on $w$ to get the runs on $w_I$, $w_J$ and $w_{I \uplus J}$, and using the commutativity of the multiplication, we get that there exist $s,s^\prime,t,t^\prime$ such that $f(w,c) = s \cdot f(w_I,c) = t \cdot f(w_J,c) = s\cdot t\cdot f(w_{I\uplus J},c)$ and $f(w,c^\prime) = s^\prime\cdot f(w_I,c^\prime) = t^\prime\cdot f(w_J,c^\prime) = s^\prime\cdot t^\prime \cdot f(w_{I\uplus J},c^\prime)$. Suppose now that we have $f(w_I,c) = f(w_I,c^\prime)$, $f(w_J,c) = f(w_J,c^\prime)$ and $f(w_{I\uplus J}, c) = f(w_{I\uplus J},c^\prime)$, we thus get that $s = s^\prime$ and $t = t^\prime$, which implies that $f(w,c) = f(w,c^\prime)$. This contradicts our assumptions. Thus, at least one of these equalities does not hold.
\end{proof}
Now, we will define some notions on "\dwroca", that will help us in the equivalence check. 

\subsection{Underlying Weighted Automata} \label{underlyingwa}

Floating runs of a "\dwroca" are isomorphic to runs of the deterministic automaton obtained by ignoring counter values. We formalise this so-called notion \emph{underlying uninitialised weighted automaton} here. 

\begin{definition}
The \emph{"underlying uninitialised weighted automaton"} of a "\dwroca" $\Autom = (Q, \Sigma,q_0, s_0, \delta_0, \delta_1, \eta_F)$ is the uninitialised deterministic weighted automaton given by $""\uwa \Autom""= (Q, \Sigma, \delta_1^\prime, \eta_F)$, where  $\delta_1^\prime$ is a transition function from $Q\times \Sigma \to Q \times \Sring$  and is defined as follows:
\[
\delta_1^\prime(q_1,a)=(q_2,s) \text{ iff } \delta_1(q_1,a)=(q_2,o,s), \text{for some } o\in \{-1,0,+1\}.
\]

\end{definition}
The automaton $"\uwa \Autom"$ is said to be \emph{uninitialised} because the machine has no initial state and weight. 
A configuration $c$ of "\dwroca" $\Autom$ is said to be $k$-equivalent to a configuration $\beta$ of an uninitialised weighted automata $\Butom$, denoted $c ""\sim""_k \beta$, if for all $w\in\Sigma^{\leq k}, f_\Autom(w,c)= f_\Butom(w, \beta)$.

Given an uninitialised weighted automaton $\Butom$ and $k > 0$, we partition the configurations of "\dwroca" $\Autom$ into two parts: $"\Dwak \Autom \Butom k"$ are those $k$-equivalent to some configuration of $\Butom$, and $"\Dcwak \Autom \Butom k"$  are those that are not.  
\AP
\begin{align*}
""\Dcwak \Autom \Butom k"" &= \{c \in Q \times \N \times \Sring\ |\ \forall \beta \in Q \times \Sring,\  c "\not\sim"_k \beta\}.\\
""\Dwak \Autom \Butom k"" &= \{c \in Q \times \N \times \Sring\ |\ \exists \beta \in Q \times \Sring,\  c "\sim"_k \beta\}.
\end{align*}

The following lemma shows that membership in $"\Dwak \Autom \Butom k"$ is independent of the weight of the configuration. We prove this by taking advantage of the existence of multiplicative inverses in the "field" and the commutativity of multiplication. 
\begin{lemma}\label{multinverse}
For all $s, \bar s \in \Sring\setminus\{0_e\},\ p\in Q,\ m\in\N$, $\configoca pms \in "\Dwak \Autom \Butom k"$ if and only if  $\configoca {p}{m}{\bar s} \in "\Dwak \Autom \Butom k"$.
\end{lemma}
\begin{proof}
Let us assume that $\configoca pms \in "\Dwak \Autom \Butom k"$. Hence, there exists a configuration $\configwa \beta t \in Q \times \Sring$  of $\Butom$ such that for all $w \in \Sigma^{\leq k}$, $f_\Autom(w, {(p,m,s)})=f_{\Butom}(w, \configwa \beta t)$.
Multiplying with $\bar s \times s^{-1}$ on both sides we get that for all $w \in \Sigma^{\leq k}$,
$
\bar s \times s^{-1} \times f_\Autom(w, {(p,m,s)})= \bar s \times s^{-1} \times f_{\Butom}(w, \configwa \beta t)
$.
Let $\bar t = \bar s \times s^{-1} \times t$. Due to the commutativity of multiplication, we get that for all $w \in \Sigma^{\leq k}$, 
$
f_\Autom(w, {(p,m,\bar s)})=  f_{\Butom}(w, \configwa \beta{\bar t})
$.
Therefore, $(p,m,\bar s) \in "\Dwak \Autom \Butom k"$. 
\end{proof}

The distance of a configuration $c$ to a set of configurations ${C}$ is the length of a minimal word that takes you from $c$ to a configuration in ${C}$ and is defined as
\[\distto(c,{C}) = \min\{|w| \mid \exists c'\in {C}\text{ with }c \xrightarrow{w} c'\}.\] 
Notice that $\distto(c,{{C}}) < \infty$ if and only if ${C}$ is reachable from $c$. We denote this by $c\rightarrow^*{{C}}$. By abuse of notation, we denote $c \xrightarrow{w} {C}$ if there exists a configuration $c'\in {C}$ such that $c \xrightarrow{w} c'$. The notion of distance will play a key role in determining which parts of the run of a non-equivalent witness can be pumped out if it is not minimal.

Consider a configuration $c$ such that $c \xrightarrow* "\Dcwak{\Autom}{\Butom}{k}"$. The following lemma identifies a special word $u$ such that $c \xrightarrow{u} "\Dcwak{\Autom}{\Butom}{k}"$. The proof of the lemma is similar to that of the non-weighted case presented in \cite{doca,droca}.

\begin{lemma} \label{specialword} \label{boundedcounter}
Given a configuration $c$ of a "\dwroca" $\Autom$, and a weighted automaton $\Butom$, if $c \xrightarrow * "\Dcwak{\Autom}{\Butom}{k}"$ then there exists a word $u=u_1u_2^ru_3$ (with $r\geq 0$) such that $c \xrightarrow{u} "\Dcwak{\Autom}{\Butom}{k}"$ and the following conditions hold:
\begin{enumerate}
\item for all $w\in \Sigma^{*}$, if  $c \xrightarrow{w} "\Dcwak{\Autom}{\Butom}{k}"$, then $|w| \geq |u|$.
\item $|u_1u_3| \leq 3|\Autom|^3$. 
\item $|u_2| \leq |\Autom|$.
\item either $u_2=\epsilon$ or $u_2$ is a simple cycle of counter loss greater than zero. 
\item \label{newpoint} $|u| \leq (max\{|\Autom|, n_c\}+|\Autom|^2)|\Autom|$ and the maximum counter value encountered during the run \\
$c \xrightarrow{u} "\Dcwak{\Autom}{\Butom}{k}"$ is less than $max\{|\Autom|, n_c\}+|\Autom|^2$. 
\end{enumerate}
\end{lemma}
\begin{proof}
Let us fix $u$ to be a minimal length word such that $\pi=c \xrightarrow{u} c^\prime \in  "\Dcwak{\Autom}{\Butom}{k}"$ for the rest of the proof.
First, we prove \Cref{newpoint}, let us assume that the maximum counter value $M$ encountered during the run $c \xrightarrow{u}  "\Dcwak{\Autom}{\Butom}{k}"$ is greater than $max\{|\Autom|, n_c\}+|\Autom|^2$. By pigeon-hole principle, we can find configurations $c_i,c_j,c^\prime_i, c^\prime_j$ and $d<|\Autom|$ such that 
$u=u_1u_2u_3u_4u_5$ and 
\[c \xrightarrow{u_1} c_i \xrightarrow{u_2} c_j \xrightarrow{u_3} c^\prime_j \xrightarrow{u_4} c^\prime_i \xrightarrow{u_5} c^\prime.
\]
where $n_{c_i}=n_{c^\prime_i}$, $n_{c_j}=n_{c^\prime_j}$, $n_{c_j}=n_{c_i}+d$, $q_{c_i}=q_{c_j}$ and $q_{c^\prime_i}=q_{c^\prime_j}$. Also, the configurations are chosen in such a way that ${c_i}$ and ${c_j}$ (resp. ${c^\prime_i}$ and ${c^\prime_j}$) are chosen to be the configurations where the counter values $n_{c_i}$ and $n_{c_j}$ (resp. $n_{c^\prime_i}$ and $n_{c^\prime_j}$) occurs for the last (resp. first) time during the run before (resp. after) reaching counter value $M$ for the last time during the run.

Now, consider the run $c \xrightarrow{u_1u_3u_5} c^{\prime\prime}=\configoca{q_{c^\prime}}{n_{c^\prime}}{s_{c^{\prime\prime}}}$. From \Cref{multinverse}, we get that $c^{\prime\prime}\in "\Dcwak{\Autom}{\Butom}{k}"$ contradicting the minimality of $u$.
We can now prove that $|u| \leq (max\{|\Autom|, n_c\}+|\Autom|^2)|\Autom|$. Assume not. By the Pigeon-hole principle, we can find configurations $c_i,c_j$  with  $q_{c_i}=q_{c_j}$ and $n_{c_i}=n_{c_j}$ such that 
$u=v_1v_2v_3$ and 
\[c \xrightarrow{v_1} c_i \xrightarrow{v_2} c_j \xrightarrow{v_3} c^\prime \in  "\Dcwak{\Autom}{\Butom}{k}".
\]
Now, consider the run $c \xrightarrow{v_1v_3} c^{\prime\prime}=\configoca{q_{c^\prime}}{n_{c^\prime}}{s_{c^{\prime\prime}}}$. Again from \Cref{multinverse}, we get that $c^{\prime\prime}\in "\Dcwak{\Autom}{\Butom}{k}"$ contradicting the minimality of $u$.

Now, we prove Items 1-4.
Assume that $|n_{c^\prime}-n_c| < |\Autom|^2$ then by \Cref{newpoint} the lemma will trivially be true with $u_2=\epsilon$.
So let us assume that $n_{c^\prime}\geq n_c+|\Autom|^2$.
First we find a simple loop $\pi^\prime= c_1\tau_1c_2 \tau_2 \ldots c_i$ in $\pi$ with the maximum "efficiency" $\varrho$. Let $v_2="word"(\pi^\prime)$ denote the corresponding word.
Now, we remove disjoint loops with maximal length from the run $\pi$ to get another run $\pi^{\prime\prime}$ such that the sum of the "counter-effect"s of the loops removed is a multiple of $\varrho$ and preserve at least one occurrence of state $q_{c_1}$ in $\pi^{\prime\prime}$. Now if $|\pi^{\prime\prime}| \geq |\Autom|^3$, then by applying the Pigeon-hole principle, we can find more loops in $\pi^{\prime\prime}$ such that the sum of their "counter-effect" is a multiple of $\varrho$ contradicting our maximality condition.

Now $\pi^{\prime\prime}= c \xrightarrow{v_1} c^\prime_1 \xrightarrow{v_2} e$
where $c^\prime_1= \configoca{q_{c_1}}{n_{c^\prime_1}}{t_{c^\prime_1}}$ and $e= \configoca{q_{c^\prime}}{n_{c^{\prime}}+r\varrho}{t_{e}}$ for some ${n_{c^\prime_1}}, r \in \N$,   ${t_{c^\prime_1}}, {t_{e}} \in \Sring$ and $v_1,v_3 \in \Sigma^*$ with $|v_1v_3|<|\Autom|^2$. Since $|n_{c^\prime}-n_c| < |\Autom|^2$ we get that $r>0$. 
Therefore, 
$c \xrightarrow{v_1} c^\prime_1\xrightarrow{v_2^r} c^\prime_2 \xrightarrow{v_3} e^{\prime}$
where $c^\prime_2= \configoca{q_{c_1}}{n_{c^\prime_1}+r\varrho}{t_{c^\prime_2}}$ and $e^\prime= \configoca{q_{c^\prime}}{n_{c^{\prime}}}{t_{e^\prime}}$ for some $t_{c^\prime_2}, t_{e^\prime} \in \Sring$.
Now, from \Cref{multinverse}, we get that $e^{\prime}\in "\Dcwak{\Autom}{\Butom}{k}"$. Note that since we have replaced loops in $\pi$ with ones having at least the same "efficiency", the length of the string $v_1v_2^rv_3$ is still minimal. 
\end{proof}

\section{Equivalence of DWROCA}
\label{sec:ptime}
\AP
In the remainder of the section, we fix two "\dwrocas", $\Autom_1$ and $\Autom_2$. We also fix $""\K"" = |\Autom_1| + |\Autom_2|$. 
To simplify the reasoning, we will reason on the synchronised runs on pairs of configurations.
Given two "\dwrocas", we consider a \emph{configuration pair} $c= \pconfig{\chi_c}{\psi_c}$ where $\chi_c$ is a configuration of $\Autom_1$ and $\psi_c$ is a configuration of $\Autom_2$.
We similarly consider \emph{transition pairs} of $\Autom_1$ and $\Autom_2$ and consider \emph{synchronised runs} as the application of a sequence of transition pairs to a configuration pair.
Let $w$ be a minimal word that distinguishes $\Autom_1$ and $\Autom_2$. Henceforth, we will denote by
\[
""\mrun"" = c_0 \tau_0 c_1 \cdots \tau_{\Ell-1} c_\Ell.
\]
the run pair of $w$ from the initial configuration pair $c_0=\confign{0}{0}{p_0}{q_0}{s_0}{t_0}$. We denote by $T_\mrun = \tau_0\cdots\tau_{\Ell-1}$ the sequence of transition pairs of this run pair. Given a run pair $\mrun'$ of a word $w'$ from $c_0$, we use $""\aw""(\mrun')$ to denote the tuple $(f_{\Autom_1}(w'),\ f_{\Autom_2}(w'))$.

\begin{lemma}\label{mainlemma}
There is a polynomial $"\Pthree":\N \rightarrow \N$ such that if two $"\dwrocas"$ $\Autom_1$ and $\Autom_2$ are not equivalent, then there exists a witness $w$ such that counter values in the execution of $w$ is less than $"\Pthree"(|\Autom_1|+|\Autom_2|)$.
\end{lemma}
This is the technically challenging part of the proof, and we will prove this later.
 Now, we show that the length of $w$ is bounded by a polynomial, assuming the counter values in $"\mrun"$ are bounded by $"\Pthree"("\K")$.

\begin{lemma}[small model property]\label{polywitness}
There is a polynomial $"\Pzero":\N \rightarrow \N$ such that for any two non-equivalent "\dwrocas" $\Autom_1$ and $\Autom_2$, the length of a "minimal witness" $w$ is less than or equal to $"\Pzero"(|\Autom_1| + |\Autom_2|)$.
\end{lemma}
\begin{proof}
Let $"\Pthree"("\K")$ be a polynomial in $"\K"$ such that the counter values encountered during the execution of $w$ are bounded by it.
Assume for contradiction that $|w|$ is greater than $""\Pzero""("\K")=2("\K""\Pthree"("\K"))^2$.
Consider the execution $"\mrun"= c_0 \tau_0 c_1 \cdots \tau_{\Ell-1} c_\Ell$ of $w$. By Pigeon-hole principle, there exist $0\leq i_1 < j_1 <i_2 <j_2\leq L$ such that if the configuration pair $c_{i_1}=\langle \configoca{p_1}{m_1}{s_1},\configoca{q_1}{n_1}{t_1} \rangle$ then $c_{j_1}=\langle \configoca{p_1}{m_1}{s^\prime_1},\configoca{q_1}{n_1}{t^\prime_1} \rangle$ and if $c_{i_2}=\langle \configoca{p_2}{m_2}{s_2},\configoca{q_2}{n_2}{t_2} \rangle$ then $c_{j_2}=\langle \configoca{p_2}{m_2}{s^\prime_2},\configoca{q_2}{n_2}{t^\prime_2} \rangle$ in $"\mrun"$.

We can deduce that $T_1 = T_{0\cdots i_1-1}T_{j_1\cdots {\Ell-1}}$, $T_2 = T_{0\cdots i_2-1}T_{j_2\cdots {\Ell-1}}$ and $T_{3} = T_{0\cdots i_1-1}$ $T_{j\cdots i_2-1} T_{j_2\cdots {\Ell-1}}$ are such that the three runs $T_1(c_0)$, $T_2(c_0)$ and $T_{3}(c_0)$ are all valid runs that are shorter than $"\mrun"$ and end in configuration pairs differing only by their weights.
Let $"\aw"("\mrun") = (s,t),\ "\aw"(T_1(c_0)) = (s_1,t_1),\ "\aw"(T_2(c_0)) = (s_2,t_2),$ and  $"\aw"(T_{3}(c_0)) = (s_3,t_3)$. 
We know from \Cref{generalTheorem} that  there exists an $i \in \{1,2,3\}$ such that $s_i \neq t_i$ 
and hence $\pi_i$ contradicts the minimality of $"\mrun"$.
\end{proof}

We now state the main result of our paper. We use a small model property stated in \Cref{polywitness} to prove this theorem. This property ensures that if two "\dwrocas" are not equivalent, then there exists a small word that can distinguish them. The small model property helps us to reduce the equivalence problem of "\dwroca" to that of weighted automata by
 ``simulating'' the runs of "\dwrocas" up to polynomial length by two weighted automata.  
\begin{theorem}
\label{equivcheck}
There is a polynomial time algorithm that decides if two "\dwrocas" are equivalent or outputs a "minimal witness". 
\end{theorem}
\label{sec:equivcheck}
\begin{proof}
Given a "\dwroca" $\Autom$ and $M\in\N$, we define the $M$-unfolding weighted automata $\Autom^M$ as a finite state weighted automaton, where the accepting weight of any word whose run does not encounter counter values greater than $M$ in $\Autom$ is equal in both $\Autom$ and $\Autom^M$.
Let $\Autom = (Q, \Sigma, q_0, s_0,\delta_0, \delta_1, \eta_F)$ be a "\dwroca" and $M\in \N$, we construct its $M$-unfolding weighted automaton 
$\Autom^M=(Q^\prime, \Sigma, p_0,s_0,\delta,\xi_{F})$ as defined below:
\begin{itemize}
    \item $Q^\prime = Q \times [0,M]$.
    \item $p_0 = (q_0,0)$.
    \item $\xi_F(q,n) =  \eta_F(q)$.
    \item $\delta((q,n),a) =  \begin{cases}
        ((q',n+o),s) \ \text{if } n= 0, \delta_0(q,a)=(q^\prime,o,s), \text{ and } n+o \leq M.\\
        ((q',n+o),s) \ \text{if } n\neq 0, \delta_1(q,a)=(q',o,s), \text{ and }n+o \leq M.\\
        \text{undefined   \hspace{.5cm} otherwise.}
        \end{cases}$
\end{itemize}

It is easy to see that for every word $w\in\Sigma^{\leq M}$, $f_\Autom(w) = f_{\Autom^M}(w)$. Furthermore $|\Autom^M| \leq |\Autom|\times M$.
We now consider two "\dwrocas" $\Autom_1$ and $\Autom_2$, and let $M = "\Pzero"(|\Autom_1|+|\Autom_2|)$ where $"\Pzero"$ is provided by \Cref{polywitness}. The lemma ensures that the two machines are not equivalent if and only if there is a word in $\Sigma^{\leq M}$ on which they differ. 

If two deterministic weighted automata (\dwa) are not equivalent, then there is a linear-sized word to distinguish them and the equivalence check can be done in polynomial time \cite[Lemma 3.4]{prob} that returns a minimal word that distinguishes them. 
Tzeng \cite{prob} shows this for two probabilistic automata. The proof can be extended to weighted automata over fields. 
Using this, we can now decide the equivalence of $\Autom^M_1$ and $\Autom^M_2$ in polynomial time and obtain a minimal word $z\in\Sigma^*$ that distinguishes them, if they are not equivalent.  We say that $\Autom_1$ and $\Autom_2$ are not equivalent if and only if $|z|\leq M$. Thus, we have a polynomial time procedure to decide the equivalence of "\dwroca", provided that \Cref{polywitness} holds.
\end{proof}
Since deterministic real-time one-counter automata is a "\dwroca" with weight $1$ on its transitions, we get the following corollary.
 \begin{corollary}
 There is a polynomial time algorithm that decides if two deterministic real-time one-counter automata are equivalent or outputs a "minimal witness".
 \end{corollary}

The rest of this section is dedicated to proving \Cref{mainlemma}.


\label{sec:polywitness}

\subsection{Proof strategy}
\label{subsec:strategy}

A crucial idea here is to partition the set of configurations between those that behave like a weighted automata on short runs and those that do not.
In \Cref{underlyingwa}, we have defined the notion of {"underlying uninitialised weighted automaton" }for a "\dwroca". Comparing the configuration of a "\dwroca" with a configuration of a weighted automaton helps us reduce the problem to that of weighted automata. The technique is adapted from the ideas used in \cite{droca,bisim} where the non-weighted model is studied. We show that the addition of weights does not change the proof outline presented in \cite{droca} and can be used for the weighted model as well with suitable adaptations.

As we need to test the equivalence of configurations from $\Autom_1$ and $\Autom_2$, we will consider the disjoint union of $"\uwa {\Autom_1}"$ and $"\uwa {\Autom_2}"$ to have a single automaton to compare their configurations with. We use $"\uwa {\Autom_1}"\cup"\uwa {\Autom_2}"$ to denote this automaton and call it the underlying weighted automata.
The set $"\NWA"$ consists of all configurations of $\Autom_1$ and $\Autom_2$ that do not have a $"\K"$-equivalent configuration in either $"\uwa{\Autom_1}"$ or $"\uwa{\Autom_2}"$ and is defined as follows
\[
""\NWA"" = \bigcup_{i \in \{1,2\}} \bigcap_{j \in \{1,2\}} "\Dcwak{\Autom_i}{\uwa{\Autom_j}}{\K}".
\]
The set $""\WA""$, the complement of $"\NWA"$, contains all configurations that are $"\K"$-equivalent to some configuration in the underlying weighted automata. Note that if a configuration $c \in "\NWA"$, then $n_c<"\K"$.
The distance that will be interesting in our reasoning is the one to the set $"\NWA"$. Therefore for simplicity, we denote $""\dist""(\xi) = \distto(\xi,{"\NWA"})$ for a configuration $\xi$ of $\Autom_1$ or $\Autom_2$.

We say that a configuration pair $c= \pconfig{\chi_c}{\psi_c}$ is
\begin{itemize}
\item[$-$] \textit{""surely-equivalent""}: If $\chi_c "\equiv"_\K \psi_c$ and $"\dist"(\chi_c) = "\dist"(\psi_c) = \infty$. 
\item[$-$] \textit{""surely-nonequivalent""}: $\chi_c "\not\equiv"_\K \psi_c$ or $"\dist"(\chi_c)\neq "\dist"(\psi_c)$. 
\item[$-$] \textit{""unresolved""}: otherwise, i.e., $\chi_c "\equiv"_\K \psi_c$ and $"\dist"(\chi_c) = "\dist"(\psi_c) < \infty$.
\end{itemize}

Let us denote by $\chi_0 = \configoca{p_0}{0}{s_0}$ and $\psi_0 = \configoca{q_0}{0}{t_0}$ the initial configurations of $\Autom_1$ and $\Autom_2$ respectively and $\chi_0 \not \equiv \psi_0$. 
The categorisation of the configuration pairs into "surely-equivalent", "surely-nonequivalent" and "unresolved" allows us to solve two easy cases before concentrating on the crux of the argument.
It is easy to observe that no configuration pair in $"\mrun"$ is "surely-equivalent" (see \Cref{surelypositive}).
Our next observation (see \Cref{surelynegative})  is that if a configuration pair $c_j$ in $"\mrun"$ is "surely-nonequivalent" and has counter values $\{m_j,n_j\}$, then the counter values in the rest of the execution is bounded by a polynomial in $m_j,n_j$ and $\K$.
Hence it will remain to prove that if $c_j$ is the first "surely-nonequivalent" configuration pair in $"\mrun"$, then $m_j$ and $n_j$ are bounded by a polynomial. In order to do  this, we prove that any minimal run-pair composed solely of "unresolved" configuration pairs has all of its counter values bounded by a polynomial. This is the most challenging part of the proof.
\[
\overbrace{c_0 \tau_0c_1\tau_1c_2 \cdots c_{j-1}\tau_{j-1} }^{\substack{\text{"unresolved" configuration pairs} \\ \text{counters poly-bounded}}} \underbrace{c_j =\pconfig{\chi_{c_j}}{\psi_{c_j}}}_{\substack{1^{st} \text{ "surely-nonequivalent"}\\ \text{ configuration pair}}} \overbrace{\tau_{j} \cdots \tau_{\Ell-1} c_\Ell}^{\substack{\text{counters} \\ \text{poly-bounded}}}
\]

We now solve the cases when we have a "surely-equivalent" or "surely-nonequivalent" configuration pair. The proof is similar to that of the non-weighted case. The case of "unresolved" configuration pairs needs to be solved using a different technique and is taken care of in \Cref{sec:unresolved}.
\begin{lemma}\label{surelypositive}
There is no \textit{"surely-equivalent"} configuration pair in $"\mrun"$.
\end{lemma}
\begin{proof}
Assume for contradiction that there exists a "surely-equivalent" configuration pair $c_i = \pconfig{\chi_{c_i}}{\psi_{c_i}}$ in $"\mrun"$. Since $"\mrun"$ is an execution of a witness $\chi_{c_i} "\not\equiv" \psi_{c_i}$. Let $v$ be a minimal witness of $\chi_{c_i}$ and $\psi_{c_i}$. Since $c_i$ is a "surely-equivalent" configuration pair, $\chi_{c_i} "\equiv"_\K \psi_{c_i}$ and thus $|v| > \K$. 
Therefore, there exists a prefix of $v$, $u \in \Sigma^{|v|-\K}$, and configurations $\chi_{c_j}$ and $\psi_{c_j}$ such that $\pconfig{\chi_{c_i}}{\psi_{c_i}} \xrightarrow u \pconfig{\chi_{c_j}}{\psi_{c_j}}$ and $\chi_{c_j} "\not\equiv"_\K \psi_{c_j}$.

 Since $v$ is a minimal witness $\chi_{c_j} "\equiv"_{\K-1} \psi_{c_j}$. Since $c_i$ is "surely-equivalent", $"\dist"(\chi_{c_i}) = "\dist"(\psi_{c_i}) = \infty$, $\chi_{c_j}$ and $\psi_{c_j}$ are in the set $"\WA"$. In other words, there exists configurations $\beta$ and $\gamma$ in the automaton $\uwa {\Autom_1}\cup\uwa {\Autom_2}$  such that $\chi_{c_j} "\sim"_\K \beta$ and $\psi_{c_j}"\sim"_\K \gamma$. Since $\chi_{c_j} "\equiv"_{\K-1} \psi_{c_j}$, it follows that $\beta "\sim"_{\K-1} \gamma$. From the equivalence result of weighted automata \cite{prob,berger}, we know that this is sufficient to prove that the automata with $\beta$ and $\gamma$ as initial distributions are equivalent, and thus, in particular, that $\beta "\sim"_\K \gamma$. That allows to deduce $\chi_{c_j} "\equiv"_\K \psi_{c_j}$, which is a contradiction.
\end{proof}

The following lemma bounds the length of a minimal witness from a "surely-nonequivalent" configuration pair.
\begin{lemma}\label{surelynegative}
Let $c_j=\pconfig{\chi_{c_j}}{\psi_{c_j}}$ be the first \textit{"surely-nonequivalent"} configuration pair in $"\mrun"$, then $\Ell-j \leq \min\{"\dist"(\chi_{c_j}),"\dist"(\psi_{c_j})\}+"\K"$.
\end{lemma}
\begin{proof}
We separately consider the two cases for $c_j$ to be "surely-nonequivalent". \\
\emph{Case-1,} $\chi_{c_j} \not\equiv_\K \psi_{c_j}$: then clearly $\Ell-j\leq "\K"$, as there is a $w\in\Sigma^{\leq "\K"}$ that is a witness from $c_j$.\\
\emph{Case-2,} $"\dist"(\chi_{c_j}) \neq "\dist"(\psi_{c_j})$: 
 Without loss of generality, we suppose $"\dist"(\chi_{c_j}) < "\dist"(\psi_{c_j})$.
    By definition of $"\dist"$, there exists a $u \in \Sigma^{\dist(\chi_{c_j})}$ such that $\chi_{c_j} \xrightarrow{u} \chi_{c_{j^\prime}}$ and $\psi_{c_j} \xrightarrow{u} \psi_{c_{j^\prime}}$ where $\chi_{c_{j^\prime}} \in "\NWA"$ and $\psi_{c_{j^\prime}} \in "\WA"$.
    By definition of $"\WA"$, there is a configuration $ \beta$ of the automaton $"\uwa {\Autom_1}"\cup"\uwa {\Autom_2}"$ such that $\psi_{c_{j^\prime}} "\sim"_\K \beta$.
    As $\chi_{c_{j^\prime}} \in "\NWA"$, $\chi_{c_{j^\prime}} "\not\sim"_\K \beta$ and thus $\chi_{c_{j^\prime}} "\not\equiv"_\K \psi_{c_{j^\prime}}$.
Therefore,
    there exists a $v \in \Sigma^{\leq \K}$ such that $ f(v,\chi_{c_{j^\prime}}) \neq f(v, \psi_{c_{j^\prime}})$ and hence $ f(uv,\chi_{c_j}) \neq f(uv, \psi_{c_j})$.
As $uv \in \Sigma^{\dist(\chi_{c_j})+\K}$, we get that $\chi_{c_j} \not\equiv_{\dist(\chi_{c_j})+\K} \psi_{c_j}$ and $\Ell-j \leq "\dist"(\psi_{c_j})+"\K"$.

In both the cases, we get that $\Ell-j \leq \min\{"\dist"(\chi_{c_j}),"\dist"(\psi_{c_j})\}+"\K"$.
\end{proof}
 From \Cref{specialword}, we know that the distance of a configuration is polynomially bounded with respect to its counter value and $"\K"$. 
Therefore, \Cref{surelynegative} tells us that the length of a minimal witness from a "surely-nonequivalent" configuration pair is polynomially bounded with respect to its counter value and $"\K"$. A length bound implies a bound on the counter value also, since the counters cannot increase more than the length of a word.
The next two sub-sections focus on the remaining case of paths containing only "unresolved" configuration pairs.

\subsection{Configuration Space}\label{subsec:partitions}

Each pair of configurations $c=\pconfig{\chi}{\psi}$ is mapped to a point in the space $\N \times \N \times (Q \times Q) \times \Sring \times \Sring$, henceforth referred to as the \textit{configuration space}, where the first two dimensions represent the two counter values, the third dimension $Q \times Q$ corresponds to the pair of control states, the fourth and fifth dimensions represent the weights. 
We partition the configuration space into "initial space", "belt space", and "background space". 
\AP This partition is indexed on two polynomials, $""\Pone""=\initialsize$ and $""\Ptwo""=\beltsize$.

\begin{itemize}
\item \emph{""initial space""}: All configuration pairs $\confign{m}{n}{p}{q}{s}{t}$ such that $m,n< "\Pone"("\K")$.
\item \emph{""belt space""}: Let $\alpha, \beta \geq 1$ be co-prime. The belt of thickness $d$ and slope $\frac{\alpha}{\beta}$ consists of those configuration pairs $\confign{m}{n}{p}{q}{s}{t}$ that satisfy $|\alpha.m - \beta.n| \leq d$. 
The "belt space" contains all configuration pairs $\confign{m}{n}{p}{q}{s}{t}$ outside the "initial space" that are inside a belt with thickness $"\Ptwo"("\K")$ and slope $\frac{\alpha}{\beta}$, where $\alpha, \beta \in [1,"\K"^2]$. 
\item \emph{""background space""}: All remaining configuration pairs. 
\end{itemize}

The projection of the configuration space onto the first two dimensions is depicted in  \Cref{fig1}.  
The polynomials $"\Pone"$ and $"\Ptwo"$ are chosen in such a way that all "unresolved" configuration pairs fall in a belt or the "initial space", and all belts are disjoint. This is proved in \Cref{beltpoint}.
In fact, these polynomials can be used as the size of the initial space and thickness of belts in \cite{droca} since the properties to be ensured are similar. 
The precise polynomials are used in the lemma proving these properties. 

\begin{figure}[htbp]
\centering
\begin{minipage}{.5\textwidth}
  \centering
    \resizebox{.77\columnwidth}{!}{%
\begin{tikzpicture}
 \draw [line width=0.3mm] [-stealth](0,0) -- (8,0)node[anchor=south, xshift=-.7cm, yshift=-.5cm] {$\N$}; 
    \draw [line width=0.3mm] [-stealth](0,0) -- (0,8) node[anchor=east, yshift=-.7cm] {$\N$};
    \draw[line width=0.3mm](0,3)--(3,3);
    \draw[line width=0.3mm](3,3)--(3,0);
    \draw[line width=0.3mm](.3,3)--(1.4,7.4);
    \draw[line width=0.3mm](1.3,3)--(2.4,7.4);
        \draw[line width=0.3mm](2.3,3)--(6.6,7.4);
    \draw[line width=0.3mm](3,2.3)--(7.4,6.8);
           \draw[line width=0.3mm](3,.3)--(7.4,1.4);
    \draw[line width=0.3mm](3,1.3)--(7.4,2.4);
     \draw (1,1.5) node[anchor=west, xshift=-.5cm]{"initial space"};
     \draw (5,3.2) node[anchor=west, xshift=-.5cm]{"background space"};
          \draw (4,4) node[anchor=west, xshift=-.5cm, rotate=45, yshift=-.3cm]{"belt space"};
               \draw (4,1.5) node[anchor=west, yshift=-.5cm, rotate=14]{"belt space"};
                    \draw (1.5,4) node[anchor=west, xshift=-.5cm, rotate=76]{"belt space"};
        \draw [densely dotted](0,0) -- (0,-.3);
        \draw [densely dotted](3,0) -- (3,-.3);
        \draw[stealth-stealth](0,-.2)--(3,-.2)node[anchor=north, xshift=-1.5cm]{\scriptsize $"\Pone"("\K")$};
          \draw[stealth-stealth](5,5.7)--(5.68,5.06)node[anchor=south, xshift=.1cm, yshift=.4cm]{\scriptsize $"\Ptwo"("\K")$};          
\end{tikzpicture}
}
\captionof{figure}{Configuration space~\cite{wodca}.}
\label{fig1}
\end{minipage}%
\begin{minipage}{.5\textwidth}
  \centering
  \resizebox{\columnwidth}{!}{%
\begin{tikzpicture}
 \draw [line width=0.3mm] [-stealth](0,0) -- (8,0)node[anchor=south, xshift=-.7cm, yshift=-.5cm] {$\N$}; 
    \draw [line width=0.3mm] [-stealth](0,0) -- (0,8.4) node[anchor=east, yshift=-.7cm] {$\N$};
    \draw[line width=0.3mm](0,3)--(3,3);
    \draw[line width=0.3mm](3,3)--(3,0);
    \draw[line width=0.3mm](.3,3)--(1.4,7.4);
    \draw[line width=0.3mm](1.3,3)--(2.4,7.4);
        \draw[line width=0.3mm](2.3,3)--(6.6,7.4);
    \draw[line width=0.3mm](3,2.3)--(7.4,6.8);
           \draw[line width=0.3mm](3,.3)--(7.4,1.4);
    \draw[line width=0.3mm](3,1.3)--(7.4,2.4);
         \draw [densely dotted](4,4) -- (5.5,4) node[anchor=west]{$\confign{m}{n}{p}{q}{s}{t}$};
        \draw [densely dotted](6,6) -- (7.5,6) node[anchor=west]{$\confign{m^\prime}{n^\prime}{p}{q}{s^\prime}{t^\prime}$};
          \filldraw (4,4) circle[radius=1.5pt] ;
            \filldraw (6,6) circle[radius=1.5pt] ;
               \draw [dotted](4,4) -- (6,6); 
\draw  plot [smooth, tension=1] coordinates {(0,0)(2,.7)(1,2)(.5,1.4)(.7,2.2)(2,2.5)(2.5,1)(2.7,2)(2.8,3)(3.2,3.4)(3.5,3.2)(4,4)};
\draw [dashed]  plot[smooth, tension=1] coordinates {(4,4) (4.3,4.4)(4.6,5.2)(5,5.4)(5.5,5.2)(6,6)};
\draw[-][gray!20](0,-.7)--(0,-.7);
\draw  plot [smooth, tension=1] coordinates {(6,6)(6.2,6.7)(7,7.4)};
\end{tikzpicture}
}
\captionof{figure}{An $\rep$ repetition.}
\label{fig2}
\end{minipage}
\end{figure}

If the maximum counter value encountered during the execution of the minimal witness is far greater than the size of the "initial space", then its length inside the "belt space" is very long. Since the belts are disjoint outside the "initial space", we just need to show that if the length of the run of the witness inside a belt is very long, then we can find a shorter witness by pumping some portion out from the current witness. 
This is proved in \Cref{sec:unresolved} using \Cref{cutlemma} and \Cref{lem:uturn}, that enables us to perform this cut.

The partition of the "unresolved" configuration pairs thus helps us to show that there is a restriction on the run containing only "unresolved" configuration pairs in $"\mrun"$.
The intuitive idea is that when at least one counter value is very big, the ratios between the counter value pairs on long run-pairs cannot take arbitrarily many different values. 
If there is a long portion of $"\mrun"$ where the ratio between the counter value pairs repeats, we can pump a sub-run out of it and still get a witness contradicting its minimality.

The proof of the following lemma is the same as that of the non-weighted case presented in \cite{droca}. 

\begin{lemma}
\label{beltpoint}
If $c=\confign{m_c}{n_c}{p_c}{q_c}{s_c}{t_c}$ is an "unresolved" configuration pair with $max\{m_c,n_c\} > "\Pone"("\K")$ then,
\begin{enumerate}
\item it lies in a unique belt of thickness $"\Ptwo"("\K")$ and slope $\rep$, where $\alpha, \beta \in [1,"\K"^2]$. 
\item it cannot take a transition pair to reach a configuration pair $c^\prime$, inside another belt of thickness $"\Ptwo"("\K")$ and slope $\frac{\alpha^\prime}{\beta^\prime}$, where $\alpha^\prime, \beta^\prime \in [1,"\K"^2]$. 
\end{enumerate}
\end{lemma}
\begin{proof}
\spnewtheorem{clam}{Claim}{\bfseries}{\itshape}
\label{sec:beltpointlemma}
To prove \Cref{beltpoint}, we first show that the following claim holds.
The proof of this claim is similar to that of the non-weighted case presented in \cite{bisim}.
\begin{clam}
\label{distancelemma}
Let $i\in\{1,2\}$ and $\chi= \configoca{q_\chi}{n_\chi}{t_\chi}$ be a configuration of $\Autom_i$. If $"\dist"(\chi)<\infty$ then, $"\dist"(\chi)= \frac{a}{b} n_\chi + d$ where $a,b \in[0, |\Autom_i|]$ and $|d| \leq 6|\Autom_i|^4$.
\end{clam}
\begin{clamproof}
Let us assume that $"\dist"(\chi)<\infty$. This means that  $\chi \rightarrow^{*} \chi^\prime$ with $\chi^\prime \in "\NWA"$. Let $k=|\Autom_i|$. Since $n_{\chi^\prime} < k$, by \Cref{specialword}, we know that there is a word $u=u_1u_2^ru_3$ (with $r\geq 0$) such that that  $\chi \xrightarrow{*} \chi^\prime$ where $|u|= "\dist"(\chi), |u_1u_3|\leq 3k^3,\ |u_2|\leq k$ and $u_2$ is a "simple cycle" with counter loss $\ell \leq k$. Let $g$ be the counter gain/loss of $u_1u_3$ and $\alpha=\frac{|u_2|}{\ell}$. Since $ |u_1u_3|\leq 3k^3$, we have $|g| \leq 3k^3$.

\begin{align*}
"\dist"(\chi)&= \frac{n_\chi- n_{\chi^\prime}- g}{\ell} |u_2| + |u_1u_3|\\
	&= \alpha n_\chi -\underbrace{\alpha(n_{\chi^\prime}+g)+ |u_1u_3|}_{d}
\end{align*}
Since $1 \leq \alpha \leq k$ it follows that $
-6 k^4 \leq d \leq 6k^4$.
\end{clamproof}
Let $c=\confign{m}{n}{p}{q}{s}{t}$ be an "unresolved" configuration pair with $max\{m_c,n_c\} > "\Pone"("\K")$. Recall that  $"\Pone"("\K") = \initialsize$ and $"\Ptwo"("\K")=\beltsize$.\\
\begin{description}
\item[1.] Let $\chi = ( p , m, s)$ and $\psi = ( q, n, t)$. Since $\pconfig {\chi}{\psi}$ is an unresolved configuration pair we have $"\dist"(\chi) = "\dist"(\psi)$. From Claim \ref{distancelemma}, there exists $a_1, b_1, a_2, b_2 \in [0,"\K"]$ and $d_1,d_2 \leq 6"\K"^4$ such that 
\[
\frac{a_1}{b_1} m + d_1 = "\dist"(\chi) = "\dist"(\psi) = \frac{a_2}{b_2} n + d_2
\]
Therefore $|\frac{a_1}{b_1} m - \frac{a_2}{b_2} n| \leq |d_2 - d_1| \leq 6"\K"^4$. This satisfies the belt condition.
\item[2.] Let $B$ and $B^\prime$ be two distinct belts with $\mu$ being the slope of belt $B$ and $\mu^\prime$ the slope of belt $B^\prime$. Hence $\mu \neq \mu^\prime$.
Without loss of generality, let us assume that $\mu^\prime > \mu$. 
It suffices to show that for all $x>\initialsize$, we have
\[
\mu x + \beltsize +1 < \mu^\prime {x} - \beltsize - 1
\]
We know that $\mu^\prime- \mu \geq \frac{1}{(\K)^2}$ and $x> \initialsize$. 
\begin{align*}
\frac{14"\K"^6}{"\K"^2}< (\mu^\prime- \mu)x \implies & \mu x +\frac{14 "\K"^4}{2} < \mu^\prime x - \frac{14 "\K"^4}{2}  \\
\implies & \mu x + \beltsize +"\K"^4 < \mu^\prime x - \beltsize - "\K"^4
\end{align*}
\end{description}
This concludes the proof.
\end{proof}

By \Cref{beltpoint}, the belt in which an "unresolved" configuration pair $c$ lies is uniquely determined. It also ensures that the belts are disjoint outside the "initial space" and that no run composed only of "unresolved" configuration pairs can go from one belt to another without passing through the "initial space".

Let $\alpha, \beta \in [1,"\K"^2]$ be co-prime and $\confign{m}{n}{p}{q}{s}{t}$, $\confign{m^\prime}{n^\prime}{p^\prime}{q^\prime}{s^\prime}{t^\prime}$ be two configuration pairs. They are  
\emph{$""\rep""$ related} if $p=p^\prime$ and $q=q^\prime$ and $\alpha.m -\beta.n = \alpha.m^\prime -\beta.n^\prime$. Roughly speaking, two configuration pairs are $"\rep"$ related
if they have the same state pairs and lie on a line with slope $\frac{\alpha}{\beta}$. 
An \emph{$"\rep"$ repetition} is a run-pair ${\bar\pi_1} =c_i \tau_i c_{i+1} \tau_{i+1} \cdots \tau_{j-1} c_j$ such that the configuration pairs $c_i$ and $c_j$ are $"\rep"$ related. The projection of an $"\rep"$ repetition onto the counters is depicted in \Cref{fig2}.

 The following two lemmas help us to show that if there is a witness having a very long sub-run inside a belt, then we can find a shorter witness whose sub-run inside that belt is shorter.

\begin{lemma}[cut lemma]\label{cutlemma} \label{repetitionbound}
    Given a configuration pair $c$ and a sequence of transition pairs $T$ such that all configuration pairs of $T(c)$ are inside a belt with slope $\frac{\alpha}{\beta}$ with $\alpha,\beta \in [1,"\K"^2]$, and either $"\ce"(T)|_1 > "\K"^2"\Ptwo"("\K")$ or $"\ce"(T)|_2 > "\K"^2"\Ptwo"(\K)$, then there exist $i < j$ such that:
    \begin{itemize}
        \item $T_{1\cdots i}(c)$ and $T_{1\cdots j}(c)$ are $"\rep"$ related configuration pairs.
        \item $T_{1\cdots i}T_{j+1 \cdots |T|}(c)$ is a run inside the same belt.
    \end{itemize}
\end{lemma}

\begin{proof}
Let us assume that $"\ce"(T)|_1 >"\K"^2"\Ptwo"("\K")$. The case for the second component can be proven analogously.
    For any $i \in [0,"\ce"(T)]$, we denote by $\ell_i$ the index such that $"\ce"(T_{0\cdots \ell_i}) = i$ and for any $k > \ell_i$, $"\ce"(T_{0\cdots k}) > i$.
    We denote $c_i = T_{0\cdots \ell_i}(c)$.
    As there are at most $"\K"^2$ many possible state pairs, and $"\Ptwo"("\K")$ many possible values for $\alpha m_c - \beta n_c$ for a configuration pair $c$ in a belt. By the Pigeon-hole principle, there exist $1\leq i < j\leq|T|$ such that $c_i$ and $c_j$ are $"\rep"$ related.
    As for every $k > \ell_j$, $"\ce"(T_{0\cdots k})> "\ce"(T_{0\cdots \ell_j})$, we have that $"\ce"(T_{\ell_j+1\cdots k}) \geq 0$. Furthermore, as $c_i$ and $c_j$ have the same control states, $T_{\ell_j+1\cdots |T|}$ can be applied to $c_i$.

    Given $k \in [\ell_j,|T|]$, we call $d_k = T_{\ell_j+1\cdots k}(c_j)$ and $e_k = T_{\ell_j+1\cdots k}(c_i)$.
    We have 
    \begin{align*}
    \alpha m_{e_k} -\beta n_{e_k} &= \alpha (m_{c_i} + "\ce"(T_{\ell_j+1\cdots k}))|_1 - \beta(m_{c_i} + "\ce"(T_{\ell_j+1\cdots k}))|_2\\
    &= \alpha m_{c_i} - \beta m_{c_i} + \alpha("\ce"({T_{\ell_j+1\cdots k}})|_1) - \beta("\ce"({T_{\ell_j+1\cdots k}})|_2).
    \end{align*}
    Also $\alpha m_{e_k} - \beta n_{e_k}
     = \alpha m_{c_j} - \beta m_{c_j} + \alpha("\ce"(T_{\ell_j+1\cdots k})|_1) - \beta("\ce"(T_{\ell_j+1\cdots k})|_2)
    = \alpha m_{d_k} - \beta n_{d_k}$, since $c_i$ and $c_j$ are $"\rep"$ related.
    Thus $d_k$ and $e_k$ are also $"\rep"$ related. As $d_k$ is inside the belt, $e_k$ is also inside the belt.
    Therefore, $T_{0\cdots \ell_i}T_{\ell_j+1\cdots |T|}(c)$ is a run inside the belt.
\end{proof}

In the following lemma, we show that if there are long runs that stay inside a belt, one can find shorter runs inside the same belt whose last configuration pairs have identical states and counter values (but not weights).

\begin{lemma}[u-turn lemma] \label{inverserep}\label{lem:uturn} 
    Let $c$ be a configuration pair, and $T$ a sequence of transition pairs such that:
    \begin{itemize}
        \item $T(c)$ is completely inside some belt with slope $\frac{\alpha}{\beta}$.
        \item $"\ce"(T) = 0$.
        \item $max("\ce"({T_{0\cdots k}})\mid 0\leq k \leq |T|) > ("\K"^2"\Ptwo"("\K")+1)^2$.
    \end{itemize}
Then, there exist $i < j < k < \ell$ such that $T' = T_{1\cdots i}T_{j+1\cdots k} T_{\ell+1\cdots |T|}$ is such that $T'(c)$ is a run inside the same belt with $"\ce"({T'}) = 0$, and the ending configuration pairs of $T'(c)$ and $T(c)$ only differ by their weights.
\end{lemma}

\begin{proof}
We denote $M = max("\ce"({T_{0\cdots k}}) \mid 0 \leq k \leq |T|)$.
For any $i \in [0,M]$, we denote by $u_i$ and $d_i$ the indices such that $"\ce"({T_{0 \cdots u_i}}) = "\ce"({T_{0 \cdots d_i}}) = i$, and for any $j \in [u_i+1,d_i-1]$, $"\ce"({T_{0 \cdots j}}) > i$. 
We call $c_i = T_{0\cdots u_i}(c)$ and $c'_i = T_{0\cdots d_i}(c)$.

By the Pigeon-hole principle, there exist $i_1<i_2 \ldots <i_{"\K"^2"\Ptwo"("\K")} \in [0,M]$ such that all configuration pairs $c_{i_r}$, $r\in[1, {"\K"^2"\Ptwo"(\K)}]$ are $"\rep"$ related. Now consider the configuration pairs $c'_{i_r}$, $r\in[1, {"\K"^2"\Ptwo"("\K")}]$. From \Cref{repetitionbound}, we know that there exist $i<j \in [1, {"\K"^2"\Ptwo"("\K")}]$ such that the configuration pairs 
$c'_i$ and $c'_j$ that are $"\rep"$ related. Note that the configuration pairs $c_i$ and $c_j$ are also $"\rep"$ related.

Similar to the previous lemma, we get that $T_{u_j+1\cdots d_j}$ is applicable to $c_i$.
Furthermore, $b_i = T_{u_j+1 \cdots d_j}(c_i)$ and $c'_i$ only differ by their weights. 
We can conclude that $T' = T_{0\cdots u_i}T_{u_j+1\cdots d_j}T_{d_i+1\cdots |T|}$ is such that $T'(c)$ is a run inside the same belt, and that $"\ce"({T'}) = 0$, and the ending states of $T$ and $T'$ are the same.
\end{proof}

\subsection{Bounding the counter values in unresolved configuration pairs}
\label{sec:unresolved}

In this section, we look at the run-pair of the minimal witness and show that counter values appearing on portions included in belts and composed only of "unresolved" configuration pairs can be bounded by some polynomial. There are two cases to consider here. The first one is where we have a very long belt visit, that does not enter the "background space", and the second is the case when it does. We use \Cref{generalTheorem}, \Cref{cutlemma} and \Cref{lem:uturn} to easily show that in the first case, the length of the minimal witness can be bounded and is proved in \Cref{uturncut}. The second and the most difficult case is when we have a very long belt visit, that enters the "background space" from the "belt space" and is considered in \Cref{lembelt}. In this case, we show that if the minimal witness reaches a "surely-nonequivalent" configuration pair whose counter values are very large, then we can pump out some portion of the run inside a belt to reach another "surely-nonequivalent" configuration pair whose counter values are polynomially bounded.

\begin{lemma}[belt lemma] \label{uturncut}
    We consider $T = \tau_0\cdots\tau_{\Ell-1}$ the sequence of transition pairs in $"\mrun"$. 
    Suppose there are $0<i<k<\Ell-1$ such that $\tau_i\cdots\tau_k(c_i)$ is included in a belt with either $m_{c_i} = m_{c_k}$ or $n_{c_i} = n_{c_k}$ then for every $i\leq r\leq k$, $"\ce"({\tau_i\cdots\tau_r})|_1 \leq 2(({"\K"^2"\Ptwo"("\K")})^2+1)$, and $"\ce"({\tau_i\cdots\tau_r})|_2 \leq 2(({"\K"^2"\Ptwo"("\K")})^2+1)$.
\end{lemma}

\begin{proof}
Let us assume that there are $0<i<k<\Ell-1$ such that $\tau_i\cdots\tau_k(c_i)$ is included in a belt with $m_{c_i} = m_{c_k}$. The case where $n_{c_i} = n_{c_k}$ can be proven analogously.
We call $R = \tau_i\cdots\tau_k$. Let $\max("\ce"({\tau_i\cdots\tau_\ell})|_1) \mid i \leq \ell \leq k) > 2(({"\K"^2"\Ptwo"("\K")})^2+1)$.
By applying \Cref{lem:uturn} twice, we can find $i_1 < j_1 \leq i_2 < j_2 < k_2 < \ell_2 \leq k_1 < \ell_1$ such that
$R_1 = R_{0\cdots i_1}R_{j_1+1\cdots k_1}R_{\ell_1+1\cdots |R|}$, $R_2 = R_{0\cdots i_2}R_{j_2+1\cdots k_2}R_{\ell_2+1\cdots |R|}$, and 
$R_{12} = R_{0\cdots i_1}R_{j_1+1\cdots i_2}R_{j_2+1\cdots k_2}R_{\ell_2+1\cdots k_1}R_{\ell_1 +1 \cdots |R|}$ are such that $R_1(c_i)$, $R_2(c_i)$ and $R_{12}(c_i)$ are runs included in the same belt as $R(c)$ and the ending configuration pairs only differ by their weights.

We can deduce that $T_1 = T_{0\cdots i-1}R_1T_{k+1\cdots |T|}$, $T_2 = T_{0\cdots i-1} R_2 T_{k+1\cdots |T|}$ and $T_{3} = T_{0\cdots i-1} R_{12} T_{k+1\cdots |T|}$ are such that the three runs $T_1(c_0)$, $T_2(c_0)$ and $T_{12}(c_0)$ are all valid runs that are shorter than $T(c)$ and end in configuration pairs only differing by their weights.

Let $"\aw"(T(c_0)) = (s,t),\ "\aw"(T_1(c_0)) = (s_1,t_1),\ "\aw"(T_2(c_0)) = (s_2,t_2)$, and $"\aw"(T_{3}(c_0))= (s_3,t_3)$. 
From \Cref{generalTheorem}, we know that  there exists an $i \in \{1,2,3\}$ such that $s_i \neq t_i$ 
and hence $T_i(c_0)$ contradicts the minimality of $"\mrun"$.
\end{proof}

\begin{figure}[htbp]
\centering

\begin{minipage}{.45\textwidth}
  \centering
  \resizebox{.448\columnwidth}{!}{%
\begin{tikzpicture}
 \draw [line width=0.4mm](.5,0) -- (2,8.1);
  \draw [line width=0.4mm](2.5,0) -- (4,8.1);
   \draw [line width=0.4mm](0,0) -- (3,0);

   \draw  plot [smooth, tension=.5] coordinates {(1,-.5) (1.2,1) (1.5,2)(1.3,2.5)};
       \draw [densely dotted](1,-.65) -- (1,-.5); 
   \draw [dashed]  plot[smooth, tension=1] coordinates { (1.3,2.5) (1.7,2.7) (1.6,2.9)(1.7,3.2)(1.9,3.5)(1.55,3.8)};
     \draw [dotted](1.3,2.5) -- (1.55,3.8); 
      \draw  plot [smooth, tension=.5] coordinates {(1.55,3.8) (1.4,4) (1.8,4.2)(2,4.5)};
      \draw [dashed] plot [smooth, tension=.5] coordinates {(2,4.5)(2.1,4.8)(1.8,5)(1.75,5.5)(2.35,6.32)};
       \draw [dotted](2,4.5) -- (2.35,6.32);

   \draw  plot [smooth, tension=.5] coordinates {(2.5,-.5)(1.7,0)(2.3,1) (2.2,2) (2.1,3)(2.3,3.5)};
    \draw [densely dotted](2.5,-.5) -- (2.7,-.6); 
   \draw [dashed]  plot[smooth, tension=1] coordinates { (2.3,3.5) (2.7,3.7) (2.8,3.9)(2.7,4.2)(2.9,4.5)(2.55,4.8)};
     \draw [dotted](2.3,3.5) -- (2.55,4.8); 
      \draw  plot [smooth, tension=.5] coordinates {(2.55,4.8) (2.4,5.2) (2.8,5.4)(3,5.5)};
      \draw [dashed] plot [smooth, tension=.5] coordinates {(3,5.5)(3.1,5.8)(3.2,6)(3.5,6.5)(3.35,7.32)};
       \draw [dotted](3,5.5) -- (3.35,7.32);
       
         \draw  plot [smooth, tension=.5] coordinates {(2.35,6.32)(3,8)(3.35,7.32)};
      
  \filldraw (1.3,2.5) circle[radius=1.5pt] node[anchor=west,xshift=0.2cm]{${i_{1}}$};
 \filldraw (1.55,3.8) circle[radius=1.5pt] node[anchor=west,xshift=0.2cm]{${j_1}$};
  \filldraw (2,4.5) circle[radius=1.5pt] node[anchor=west,xshift=0cm]{${i_2}$};
 \filldraw (2.35,6.32) circle[radius=1.5pt] node[anchor=west,xshift=0cm]{$j_2$};
 
  \filldraw (2.3,3.5) circle[radius=1.5pt] node[anchor=west,xshift=0.2cm]{${l_{1}}$};
 \filldraw (2.55,4.8) circle[radius=1.5pt] node[anchor=west,xshift=0.2cm]{${k_1}$};
  \filldraw (3,5.5) circle[radius=1.5pt] node[anchor=west,xshift=0cm]{${l_2}$};
 \filldraw (3.35,7.32) circle[radius=1.5pt] node[anchor=west,xshift=0cm]{$k_2$};
\end{tikzpicture}
}
\captionof{figure}{A belt visit returning to the "initial space"}
\label{belt1}
\end{minipage}%
\begin{minipage}{.55\textwidth}
\centering
\resizebox{.652\columnwidth}{!}{%
\begin{tikzpicture}
 \draw [line width=0.4mm](1,2.5) -- (7,8.5);
  \draw [line width=0.4mm](1.5,0) -- (8,6.5);
\draw  plot [smooth, tension=.5] coordinates {(1.5,1.5) (2,1.5) (2.5,2.5)};
\draw [dashed]  plot[smooth, tension=1] coordinates { (2.5,2.5) (3,3.5)(3.5,2.6) (4,4)};
\draw  plot [smooth, tension=1] coordinates {(4,4) (4.6,5.3)(5,5.4)};
\draw [dashed]  plot[smooth, tension=1] coordinates {(5,5.4)(5.5,5.2)(6.2,6.6)};
\draw  plot [smooth, tension=1] coordinates {(6.2,6.6) (6.4,7.1)(7,7.7)};
  \filldraw (1.5,1.5) circle[radius=1.5pt] node[anchor=east,xshift=-0.2cm] {${i}$};
  \filldraw (2.5,2.5) circle[radius=1.5pt] node[anchor=north,xshift=0.2cm]{${i_{1}}$};
 \filldraw (4,4) circle[radius=1.5pt] node[anchor=north,xshift=0.2cm]{${j_1}$};
  \filldraw (5,5.4) circle[radius=1.5pt] node[anchor=north,xshift=0.2cm]{${i_2}$};
 \filldraw (6.2,6.6) circle[radius=1.5pt] node[anchor=north,xshift=0.2cm]{$j_2$};
  \filldraw (7,7.7) circle[radius=1.5pt] node[anchor=west,xshift=0.2cm, yshift=0.2cm]{$j$};
 \draw [dotted](5,5.4) -- (6.2,6.6);
  \draw [dotted](2.5,2.5) -- (4,4);
\end{tikzpicture}
}
\captionof{figure}{A belt visit ending in a belt}
\label{belt2}
\end{minipage}
\end{figure}

\begin{lemma} 
\label{lembelt}
Let $c_j$ be the first {"surely-nonequivalent"} configuration pair in $"\mrun"$. Then $max\{m_{c_j}, n_{c_j}\}$ $\leq "\Pone"("\K")+"\K"({"\K"^2"\Ptwo"("\K")})+1$.
\end{lemma}
\begin{proof}
Assume for contradiction that the execution of a minimal witness $w$ reaches a "surely-nonequivalent" pair $c_j=\confign{m_{c_j}}{n_{c_j}}{p_{c_j}}{q_{c_j}}{s_{c_j}}{t_{c_j}}$, with $max\{m_{c_j}, n_{c_j}\}$ greater than $"\Pone"("\K")+"\K"({"\K"^2"\Ptwo"("\K")})+1$. Since $c_j$ is the first "surely-nonequivalent" configuration pair during the execution of a minimal witness, we know that all the previous configuration pairs are "unresolved" configuration pairs and lie either in the "initial space" or "belt space". The idea here is to find an $"\rep"$ repetition that is a factor of the execution inside a belt that can be cut out to obtain a shorter witness (refer \Cref{belt2}).

Let $T = \tau_0\cdots\tau_{\Ell-1}$ be the sequence of transition pairs corresponding to a minimal witness from $c_0$.  For all $0<i\leq\Ell$, let $c_i= \tau_0\cdots\tau_{i-1}(c_0)$. We use $c_j= \pconfig{\chi_{c_j}}{\psi_{c_j}}$, $j\leq \Ell$, to denote the first "surely-nonequivalent" configuration pair during the execution of $w$, where $\chi_{c_j}=\configoca {p_{c_j}}{m_{c_j}}{s_{c_j}}$ and $\psi_{c_j}=\configoca {q_{c_j}}{n_{c_j}}{t_{c_j}}$.
Let $T_1 = \tau_0\cdots\tau_{j-1}$ and $w=v_1v_2$, where $v_1="word"(T_{0\cdots j-1})$ and $v_2="word"(T_{j\cdots\Ell-1})$. Since $c_j$ is "surely-nonequivalent" either $(1)$ $\chi_{c_j} "\not\equiv"_\K \psi_{c_j}$ or $(2)$ $"\dist"(\chi_{c_j})\neq "\dist"(\psi_{c_j})$.

\begin{flushleft}
\emph{\textbf{Case-1:}} $\chi_{c_j} \not\equiv_\K \psi_{c_j}$.\\
\end{flushleft}
Since $\chi_{c_j} "\not\equiv"_\K \psi_{c_j}$, we know that $|v_2| \leq "\K"$. Also, since $\max\{m_j, n_j\}$ is greater than $"\Pone"("\K")+2({"\K"^2"\Ptwo"("\K")}+1)$,
by applying \Cref{repetitionbound} twice, we can find $i<i_1 < j_1 \leq i_2 < j_2 < j$ such that $R(c_i)=T_{i \cdots j}(c_i)$ is a run inside a belt and
$R_1 = T_{i\cdots i_1}T_{j_1+1\cdots j}$, $R_2 = T_{i\cdots i_2}T_{j_2+1\cdots j}$, and 
$R_{12} = T_{i\cdots i_1}T_{j_1+1\cdots i_2}T_{j_2+1\cdots j}$ are such that $R_1(c_i)$, $R_2(c_i)$ and $R_{12}(c_i)$ are runs inside in the same belt as $R(c_i)$ as shown in \Cref{belt2}.

We can deduce that $T_1 = T_{0\cdots i-1}R_1T_{j+1\cdots |T|}$, $T_2 = T_{0\cdots i-1} R_2 T_{j+1\cdots |T|}$ and $T_{3} = T_{0\cdots i-1} R_{12} T_{j+1\cdots |T|}$ are such that the three runs $T_1(c_0)$, $T_2(c_0)$ and $T_{12}(c_0)$ are all valid runs that are shorter than $T(c)$ and end in configuration pairs having the same states.
Let $"\aw"(T(c_0)) = (s,t),\ "\aw"(T_1(c_0)) = (s_1,t_1),\ "\aw"(T_2(c_0)) = (s_2,t_2)$, and  $"\aw"(T_{3}(c_0)) = (s_3,t_3)$. 
We know from \Cref{generalTheorem} that  there exists an $i \in \{1,2,3\}$ such that $s_i \neq t_i$ 
and hence $T_i(c_0)$ contradicts the minimality of $"\mrun"$.

\clearpage
\begin{flushleft}
\emph{\textbf{Case-2:}}  If $"\dist"(\chi_{c_j})\neq "\dist"(\psi_{c_j})$.\\
\end{flushleft}
We show that if $\max\{m_{c_j},n_{c_j}\} > "\Pone"("\K")+"\K"("\K"^2"\Ptwo"("\K")) +1$, then we can cut out some sub-runs to get a shorter witness contradicting the minimality of $w$.

Without loss of generality  assume  ${"\dist"}(\chi_{c_j})< {"\dist"}(\psi_{c_j})$ and $\max\{m_{c_j},n_{c_j}\} > "\Pone"("\K")+"\K"("\K"^2"\Ptwo"("\K")) +1$. Since $c_j$ is the first "surely-nonequivalent" configuration pair in $"\mrun"$, for all $i\in[0,j-1]$, the configuration pairs $c_i$ are either in the "initial space" or inside a belt. Let $k$ be the smallest index such that for all $i\in[k,j-1]$, the configuration pairs $c_i$ are inside a fixed belt. Since $c_k$ is the first point inside a belt, $max(m_{c_k},n_{c_k})= "\Pone"("\K")+1$. 

Consider the run  $\tau_k\cdots\tau_j(c_k)$. By the Pigeon-hole principle, there is a set of $"\K"+1$ configuration pairs that are $"\rep"$-related to each other. Let $d_{0}, d_{1}, \ldots , d_{\K}$ denote these configuration pairs such that $m_{d_i}< m_{d_j}$ for $i<j$. 
Since ${"\dist"}(\chi_{c_j})<\infty$, we know that there exists a word $u$ such that $ \chi_{c_j} \xrightarrow{u} (p^\prime,m^\prime,s^\prime) \in "\NWA"$ with $m<"\K"$ and $|u|="\dist"(\chi_{c_j})$. Let $\chi_{0}, \chi_{1},\ldots, \chi_{\K}$ denote the configurations, where counter values 
$m_{d_{0}}, m_{d_{1}}, \ldots m_{d_{\K}}$ are encountered for the first time during the run $ \chi_{c_j} \xrightarrow{u} (p^\prime,m^\prime,s^\prime)$.
By the Pigeon-hole principle, there exist $r<l$, such that $\chi_{l}$ and $\chi_{r}$ have the same state. 
Let $u=u_1u_2u_3$ for some $u_1,u_2,u_3\in\Sigma^*$ such that $\chi_{c_j}\xrightarrow{u_1}\chi_{l} \xrightarrow{u_2}\chi{r}\xrightarrow{u_3}(p^\prime,m^\prime,s^\prime) $ and $t$ denote the difference in counter values between $\chi_{l}$ and $\chi{r}$. Since  $\chi_{r}$ and $\chi_{l}$ have the same state, for any $s \in \Sring$, the run $(p_{c_j},m_{c_j}-t,s)\xrightarrow{u_1u_3}(p,m,s^{\prime\prime})$ is a valid run for some $s^{\prime\prime}\in \Sring$.
Therefore, for all $s\in\Sring$,  $"\dist"(p_{c_j},m_{c_j}-t,s)<\infty$.

Note that $t=m_{d_l}-m_{d_r}$, is the same as the difference in counter values between $\chi_{l}$ and $\chi{r}$. Since $d_l$ and $d_r$ are $"\rep"$-related configuration pairs, removing the sub-run between them from the run $\tau_k\cdots\tau_j(c_k)$ will take us to the "background space" point  $\pconfig{\chi^\prime}{\psi^\prime}$, where $\chi^\prime= (p_{c_j},m_{c_j}-t,s)$, for some $s\in\Sring$.   Since $\pconfig{\chi^\prime}{\psi^\prime}$ is a point in the "background space" either $"\dist"(\chi^\prime)="\dist"(\psi^\prime)=\infty$ or $"\dist"(\chi^\prime) \neq "\dist"(\psi^\prime)$. We already know that for all $s\in\Sring$, $"\dist"(p_{c_j},m_{c_j}-t,s)< \infty$, therefore $"\dist"(\chi^\prime) \neq "\dist"(\psi^\prime)$. Hence, $c_{j^\prime}=\pconfig{\chi^\prime}{\psi^\prime}$ is a "surely-nonequivalent" configuration pair with $max\{m_{c_{j^\prime}}, n_{c_{j^\prime}}\}\leq "\Pone"("\K")+"\K"("\K"^2"\Ptwo"("\K"))+1$ contradicting our initial assumption.
 \end{proof}

 Note that the technique used for proving \emph{{Case-2}} is different from the one used for deterministic real-time one-counter automata by B\"ohm et al.~\cite{droca}. Using a similar technique as mentioned above will help get a better polynomial bound on the maximum counter value encountered during the run of the minimal witness while checking the equivalence of two deterministic real-time one-counter automata.

We finally conclude the proof of our main lemma.
 
\textbf{Proof of \Cref{mainlemma}. }
Let  $"\mrun"=c_0\sigma_0c_1\sigma_1\dots c_\Ell$ is the run of a minimal witness. From  \Cref{surelypositive}, we know that there is no "surely-equivalent" configuration pair in it. Let us assume that $c_j=\pconfig{\chi_{c_j}}{\psi_{c_j}}$ is the first "surely-nonequivalent" configuration pair in $"\mrun"$. 
\Cref{surelynegative} ensures that $\Ell-j \leq \mathrm{min}("\dist"(\chi_{c_j}),"\dist"(\psi_{c_j})) + \K$, therefore all counter values appearing at position greater than $j$ are bounded by $\mathrm{max}(n_{\chi_{c_j}},n_{\psi_{c_j}}) + \mathrm{min}("\dist"(\chi_{c_j},"\dist"(\psi_{c_j}))) + "\K"$.

\Cref{uturncut} ensures that all counter values appearing before $c_j$ inside belts are bounded by the value $"\Pone"("\K")+2(({"\K"^2"\Ptwo"("\K")})^2+1)$, and by \Cref{lembelt}, $\mathrm{max}(n_{\chi_{c_j}},n_{\psi_{c_j}}) \leq "\Pone"("\K")+"\K"({"\K"^2"\Ptwo"("\K")})+1$.

\AP
Finally, \Cref{specialword} ensures that $"\dist"(\chi_{c_j})$ and $"\dist"(\psi_{c_j})$ cannot be more than $(max\{"\K", n_{\chi_{c_j}}, m_{\psi_{c_j}}\}+"\K"^2)"\K"$. 
Therefore, we conclude that there exists a polynomial $""\Pthree""="\Pone"("\K")+2({"\K"^2"\Ptwo"("\K")})^2+1)$ such that all counter values appearing in $"\mrun"$ are bounded by $"\Pthree"("\K")$. 
\qed

From \Cref{mainlemma}, we get that the value of $"\Pthree"$ is $\mathcal{O}("\K"^{12})$ and from \Cref{polywitness}, we get that the value of $"\Pzero"$ is $\mathcal{O}(\K^{26})$. 
Hence, if two "\dwrocas" $\Autom$ and $\Butom$ are not equivalent, then there is a word whose length is of $\mathcal{O}("\K"^{26})$ such that $f_\Autom(w)\neq f_\Butom(w)$.

\clearpage

\section{Conclusion and Future work}
\label{sec:conclusion}

In this paper, we presented an improved polynomial time algorithm for checking the equivalence of two "\dwrocas" that return a word distinguishing them, if it exists. 
A potential research direction is to remove the ``real-time'' constraint in "\dwroca". Note that equivalence for deterministic one-counter automata is $\CF{NL}$-complete regardless of whether it is real-time. However, the techniques for the non-real-time case differ from those used here. Therefore, it's not guaranteed that those techniques are robust enough to the addition of weights, which makes it an interesting topic for further investigation. 
Finally, a polynomial time algorithm for equivalence of "\dwroca" also paves the way for efficient learning algorithms for "\dwroca". 
%
%
%

\begin{credits}
\subsubsection{\ackname} A.V. Sreejith would like to acknowledge the support by SERB for the project ``Probabilistic Pushdown Automata'' [MTR/2021/000788].
\end{credits}

\bibliographystyle{splncs04}
\bibliography{paper}

\ 
\end{document}